\documentclass{article}                    
\usepackage{graphicx}
\usepackage{amsmath}
\begin{document}

\title{The reaction-free trajectories \\of a classical point charge}%
\author{Michael Ibison \\ \\
             Institute for Advanced Studies at Austin \\
             11855 Research Boulevard, Austin, TX 78759, USA\\
             ibison@ias-austin.org
}

\maketitle

\begin{abstract}
It is well-known that a classical point charge in 1+1 D hyperbolic motion in space and time is reaction-free. But this is a special case of a larger set of reaction-free trajectories that in general are curved paths through space, i.e. in 2+1 D. This note catalogs the full family of reaction-free trajectories, giving a geometrical interpretation by which means the curved path possibility is easily related to the better known case of hyperbolic motion in 1+1 D. Motivated by the geometry, it is shown how the catalog of motions can be naturally extended to include the possibility of lossless reaction-free closed spatial orbits that turn out to be classical pair creation and destruction events.

The extended theory can accommodate a vacuum plenum of classical current that could be regarded as a classical version of the Fermionic ZPF of QFT, reminiscent of the relationship between the Electromagnetic ZPF and the classical imitation that characterizes `Stochastic Electrodynamics'.

keywords: Lorentz-Dirac; Abraham; von Laue; radiation-reaction; superluminal; tachyon; ZPF


\end{abstract}

\section{The Abraham - von Laue vector} \label{sec:AVLvector}
\subsection{Background} \label{subsec:background}
In Heaviside units with $c = 1$, the Lorentz-Abraham-Dirac (henceforth LAD) equation for a classical charge of mass $m$ is%
\begin{equation}  \label{1}
m a = f_{ext} + \Gamma
\end{equation}
where $ f_{ext}$ is the external force 4-vector, and $\Gamma$ is the von Laue - or Abraham  -  4-vector given by
\begin{equation} \label{2}
\Gamma  =   \frac{{2e^2 }} {3}\left( {\frac{{da}} {{d\tau }} + a^2 u} \right)\,,
\end{equation}
where $a$ and $u$ are, respectively, the proper acceleration and proper velocity 4-vectors and where - introducing the Lorentz scalar product symbol and metric used hereafter - $ a^2  \equiv a \circ a = a^\mu  a_\mu   = a_0 ^2  - {\mathbf{a}}^2 $.\footnote{We include Abraham in the authorship of (\ref{1}) in accord with the position taken by Rohrlich \cite{RohrlichAmJoP}.} The first of the two terms in $ \Gamma $ is the Schott term - also called the `acceleration reaction force' by von Laue. The second term is called the radiation reaction \cite{RohrlichAmJoP}. Noting that
\begin{equation} \label{3}
\frac{d} {{d\tau }}\left( {u \circ a} \right) = 0 = a^2  + u \circ \frac{{da}} {{d\tau }} \,,
\end{equation}
the Abraham - von Laue vector can also be written
\begin{equation} \label{4}
\Gamma   =   \frac{{2e^2 }} {3}P\circ\frac{{da}} {{d\tau }}
\end{equation}
where
\begin{equation} \label{eqn:P}
P^\mu  {\kern 1pt} _\nu \equiv 1 - u^\mu  u_\nu\,.
\end{equation}
Here and henceforth for any tensor $T$ and vector $v$: $\left\{T\circ v\right\}^\mu = T^{\mu}{\kern 1pt}_{\nu} v^{\mu}$. From the form (\ref{4}) it is clear that $ u \circ \Gamma  = 0 $, a result demanded of any four vector supplement to the relativistic equation of motion $ m_e a =   f_{ext} $, because, if for example the external force is the Lorentz force, then one already has $ u \circ a = u \circ f_{ext} = 0$.

The Abraham - von Laue vector can be derived from the action of the retarded EM fields of a charged sphere upon itself
in the limit that the radius of the sphere goes to zero \cite{RohrlichSelfForce,Erber}. Obviously, though they are responsible
for singular self-energy, the retarded fields of a uniformly moving charge can produce no self-force, from which it
follows that the Abraham - von Laue vector stands for the retarded self-force of a charge in \emph{non-uniform} motion.
It is to be noted that because the self-force can be temporarily non-zero even when the proper acceleration is zero - $
a^2 u = 0 $, $ da/d\tau  \ne 0 $ - non-uniform motion responsible for a non-vanishing Abraham - von Laue vector does not exclusively imply acceleration. By contrast, in the \emph{reaction-free} case under consideration here, the self-force
is zero even though the charge is accelerating: $ \Gamma  = 0$, $ a^2 \ne 0 $.

\subsection{Radiation} \label{subsec:radiation}
Historically it was held that $ \Gamma  = 0 $ implies that the charge is not radiating \cite{Born,Pauli}. But later analysis \cite{Bondi,Fulton,Boulware,Bradbury} and commentary \cite{Drukey,Erber} decided in favor of the presence of radiation whenever there is acceleration - independent of the value of $ \Gamma $. In this case the Lorentz-invariant generalization of the radiated power is given by the relativistic Lamor formula, $ P = - 2e^2 a^2 /3 $, and if $ \Gamma  = 0 $ then the charge produces radiation with no net reaction back upon the source. A more recent exchange \cite{Singal1,Singal2,Parrott1,Parrott2}, however, has re-opened the issue, turning on the definition of uniform acceleration `for all time'. In that case, in order to maintain that $ \Gamma  = 0 $ implies no radiation, one would have to modify the relativistic Lamor formula somehow, a non-relativistic example of which has been given by
Peierls \cite{Peierls}.

The Abraham - von Laue vector is at first a mathematical entity - expressed in terms of the motion of the electron. It has the meaning of a force by virtue of its form and placement in the Lorentz-Dirac equation. And its vanishing or otherwise is decided by the electron motion without regard to the physical interpretation. More specifically the value of the vector can be decided without regard to the emission of radiation. Since therefore it is not necessary for the purposes of the analysis in this paper, here we do not take a position on this controversial issue. Even so the discussion is re-opened in Section \ref{sub:RadiationAgain}, where it is pointed out that the relationship between radiation and the Abraham - von Laue force must be considered afresh in the geometrical extension to the superluminal domain.

\subsection{Elimination of runaway behavior} \label{subsec:runaways}
Incorporation of the Abraham - von Laue force in the relativistic Newton equation of motion gives rise to the possibility of runaway solutions to (\ref{1}). This is most easily demonstrated when there is no external force and in one space dimension, whereupon the substitution $u_x=\sinh\left(w\left(\tau\right)\right)$ gives
\begin{equation} \label{4.5}
\frac{dw} {{d\tau}} = \tau_e\frac{d^2w} {{d\tau^2}}
\end{equation}
where $\tau_e \equiv 2e^2/3m$. This has the (runaway) solution $w\sim \exp(\tau/\tau_e)$. Of course, the runaway tendency remains present even when the external force does not vanish. The traditional remedy is to require that the acceleration vanish in the distant future, presuming all external forces are zero there also. (Parrott  \cite{Parrottbook} presents an interesting and more thorough discussion of the possibilities.) That requirement can be imposed by integrating (\ref{1}) thereby converting it to a integro-differential equation, confering an opportunity to impose the desired boundary condition $a\left(\infty\right) = 0$. This is easily achieved in 1+1 D whereupon, if $f_{ext}$ is the Lorentz force, one obtains
\begin{equation} \label{eqn:1plus1D}
  m\frac{{d^2 {\mathbf{x}}}}
{{d\tau ^2 }} =  e\gamma \int\limits_0^\infty  {ds} e^{ - s} {\mathbf{E}}\left( {\tau  + \tau _e s} \right)
   \Rightarrow m\frac{{dw}}
{{d\tau }} =  e\int\limits_0^\infty  {ds} e^{ - s} E_x \left( {\tau  + \tau _e s} \right)\,.
\end{equation}
In the more general case Rohrlich \cite{RohrlichBook} tried unsuccessfully to impose the future boundary condition by straight-forward integration of the LAD equation in 3+1 D. Relative to the long history of the Lorentz-Dirac equation the full solution to that problem was found only recently \cite{Ibison1}:
\begin{equation} \label{eqn:MILAD}
ma\left( \tau  \right) = R^{ - 1} \left( \tau  \right)\int\limits_0^\infty  {ds} e^{ - s} R\left( {\tau  + s\tau_e } \right)f_{ext} \left( {\tau  + s\tau_e } \right)
\end{equation}
where
\begin{equation} \label{eqn:R}
R = \left\{ {R^{\mu} {\kern 1pt}_{\nu}  } \right\} = \left( {\begin{array}{*{20}c}
   {u_0 } & {u_1 } & {u_2 } & {u_3 }  \\
   {u_1 } & {u_0 } & 0 & 0  \\
   {u_2 } & 0 & {u_0 } & 0  \\
   {u_3 } & 0 & 0 & {u_0 }  \\
\end{array} } \right)\,.
\end{equation}
Both (\ref{eqn:1plus1D}) and (\ref{eqn:MILAD}) satisfy the respective LAD equations in 1+1 D and 3+1 D and are free from runaway behavior.

There is not a universal agreement on the correctness of the traditional Abraham - von Laue vector in the classical theory - apart from disagreements on the presence of radiation. It seems likely that interest in alternatives to the LAD equation were driven at least in part by the failure to integrate the full 3+1 D version of the equation, with the consequent failure to eliminate the runaway behavior in the more general case.

Ford and O'Connell \cite{FordEtAl} give an approximate solution to the non-relativistic LAD in 3+1 D that is free from runaways and which is cited by Jackson \cite{Jackson} as a `sensible alternative' to the original equation in that domain. Spohn \cite{Spohn} gives a similar result valid in the relativistic domain and which turns out to be equivalent to the truncation of a series expansion of (\ref{eqn:MILAD}). Specifically, using that $f_{ext} = F\circ u,\quad\left(f_{ext}^{\mu} =F^\mu  {\kern 1pt} _\nu u^\nu\right) $, (\ref{eqn:MILAD}) can be written
\begin{equation} \label{eqn:tauSeries}
ma = e R^{ - 1} \sum\limits_{n = 0}^\infty  {\left( {\tau _e \frac{d}
{{d\tau }}} \right)} ^n  R F\circ u
\end{equation}
which, to second order in $\tau_e$ is
\begin{equation} \label{eqn:twoTerms1}
\begin{aligned}
ma = & eF\circ u+ e\tau _e \left(\frac{dF}{d\tau} \circ u +F\circ a + R^{ - 1}\left(u\right)\frac{dR\left(u\right)}{d\tau}F \circ u\right) +O(\tau_e^2) \\
= & e\left(F+\tau _e\frac{dF}{d\tau} +\tau_e\frac{e}{m} F^2+ \tau_e\frac{e}{m}R^{-1}\left(u\right)R\left(F \circ u\right)F\right)  \circ u +O(\tau_e^2) \,.\\
\end{aligned}
\end{equation}
Using that the inverse to $R$ is
\begin{equation} \label{eqn:Rinv}
R^{ - 1}  = \left( {\begin{array}{*{20}c}
   {u_0 } & { - u_1 } & { - u_2 } & { - u_3 }  \\
   { - u_1 } & {\left( {1 + u_1 ^2 } \right)/u_0 } & {u_1 u_2 /u_0 } & {u_1 u_3 /u_0 }  \\
   { - u_2 } & {u_1 u_2 /u_0 } & {\left( {1 + u_2 ^2 } \right)/u_0 } & {u_2 u_3 /u_0 }  \\
   { - u_3 } & {u_1 u_3 /u_0 } & {u_2 u_3 /u_0 } & {\left( {1 + u_3 ^2 } \right)/u_0 }  \\
 \end{array} } \right)\,,
\end{equation}
and that $F^{\mu\nu}$ is antisymmetric, (\ref{eqn:twoTerms1}) simplifies to
\begin{equation} \label{eqn:twoTerms2}
ma = e\left(F+\tau _e\frac{dF}{d\tau} +\tau_e\frac{e}{m} P F^2\right)\circ u +O(\tau_e^2)\,.
\end{equation}
Eq. (\ref{eqn:twoTerms2}) with second and higher order terms in $\tau_e$ set to zero is t,he result given by Spohn. When there is no force the acceleration is zero, so in this limit (\ref{eqn:twoTerms2}) inherits from (\ref{eqn:tauSeries}) and (\ref{eqn:MILAD}) the property that they are free of runaways. Spohn's result was subsequently endorsed by Rohrlich \cite{RohrlichCorrectEquation}, which the assertion that it is "the correct equation of motion" for the classical electron.

However, the presence or absence of the omitted terms of the series is of interest in determination of the limits of the classical theory and the correct correspondence with QM. In addressing the latter Moniz and Sharp \cite{MonizSharp} found that the non-relativistic Heisenburg equations of motion generate an infinite series similar to (\ref{eqn:tauSeries}), though corresponding to a classical electron of radial dimension equal to the Compton wavelength. Barut and Zanghi \cite{BarutZanghi} investigated the classical QM correspondence through a sympletic system designed to give the Dirac equation upon quantization. They found the relativistic Heisenburg equations of motion contain a third order derivative term similar to, but not identical with, the Abraham - von Laue vector in the classical LAD. Both of these findings seem to suggest that whilst the truncated series (\ref{eqn:twoTerms2}) no doubt has merit as a useful approximation, the original third order equation - (\ref{1}) with (\ref{2}) - and correspondingly the full series (\ref{eqn:tauSeries}), are more fundamental.

In their analysis Moniz and Sharp \cite{MonizSharp} found that neither the quantum theory nor the classical series expansion for the equation of motion suffered from runaway behavior. They attribute the lack of runaways in the classical case to the finite form factor used to generate the series. According to the analysis in \cite{Ibison1} however, the series (\ref{eqn:tauSeries}) corresponding to a classical point particle is unconditionally free of runaway behavior. Even so, the analysis of \cite{Medina} supports the claim that runaway behavior can be eliminated at the level of the LAD equation if the source of the self-action that generates the Abraham - von Laue 4 vector has a finite size.  Further support comes from a paper by Rohrlich \cite{RohrlichSelfForce} who points out that if the electron is modeled as a finite charged spherical shell then the retarded self-action generates a difference relation in place of the third derivative
\begin{equation} \label{eqn:difference}
\Gamma \left(\tau \right)  =   \frac{e^2 } {3a^2}P\left(\tau \right)\circ u\left(\tau-2a \right)
\end{equation}
so that the total LAD equation becomes a second-order differential difference equation. Here $a$ is the radius of the charged sphere and $P$ is given by (\ref{eqn:P}). Non-linear terms in the derivatives of the acceleration have been ignored. Rohrlich claims the resulting LAD equation is free of `pathological solutions'.

\section{The reaction-free trajectories} \label{sec:ReactionFreeTrajectories}
The reaction-free condition that there be no retarded self-force acting back upon the charge is that the Abraham - von Laue vector is zero:\footnote{Here and throughout, by `reaction-free' is meant free of a force from the Abraham von Laue vector. It is not meant to imply the absence of radiation, which may or may not coincide with this condition, according to the dynamics and the position held on the issue of correspondence between these two.}
\begin{equation} \label{5}
\frac{{da}} {{d\tau }} + a^2 u = 0 \,.
\end{equation}
Because $ a \circ u = 0 $ the Lorentz scalar product with the acceleration gives
\begin{equation} \label{6}
a \circ \frac{{da}} {{d\tau }} = 0 \Rightarrow a^2  =  - k^2
\end{equation}
where $k$ is a real constant. The sign follows because $a$ is space-like, which follows in turn from $ a \circ u = 0 $ and that $u$ is time-like. With this (\ref{5}) implies that
\begin{equation} \label{7}
\frac{{d^2 u}} {{d\tau ^2 }} = k^2 u \,,
\end{equation}
the general solution of which can be written
\begin{equation} \label{8}
x = r + \left( {p\cosh \left( {k\tau } \right) + q\sinh \left( {k\tau } \right)} \right)/k
\end{equation}
where $ p $, $ q $ and $ r $ are constant 4-vectors, which is the result given by Rohrlich \cite{Rohrlich1}. $ p $ and $ q $ are not entirely
arbitrary, but must be chosen to satisfy (\ref{6}):
\begin{equation} \label{9}
\left( {p\cosh \left( {k\tau } \right) + q\sinh \left( {k\tau } \right)} \right)^2  =  - 1\,.
\end{equation}
Since this must be true for all $t$ it follows that
\begin{equation} \label{10}
p^2  =  - 1,\quad q^2  = 1,\quad p \circ q = 0\,.
\end{equation}
With these, (\ref{8}) then gives that $ u^2  = 1 $ , as required. Eq. (\ref{8}) with Eq. (\ref{10}) gives the full family of reaction-free trajectories for the classical charged particle.

Let us choose the origin of $\tau$ so that $ \tau = 0 $ at $ t = r_0 $. Then (\ref{8}) gives that $ p_0  = 0 $, whereupon (10) gives
\begin{equation} \label{11}
p = \left( {0,{\mathbf{\hat p}}} \right),\quad q = \left( {\sqrt {1 + {\mathbf{q}}^2 } ,{\mathbf{q}}} \right),\quad {\mathbf{p}}{\mathbf{.q}} = 0
\end{equation}
where the sign of $ q_0 $ has been chosen so that $\tau$ is an increasing function of $t$.

Since $\bf{p}$ and $\bf{q}$ are orthogonal, it is convenient to suppose that the space axes have been oriented so that $\bf{p}$ and $\bf{q}$ are parallel to the $ x_1 $ and $ x_2 $ axes respectively. Let us suppose also that the space and time axes have been located so that $r = 0$. Then (\ref{8}) becomes
\begin{equation} \label{12}
x   \equiv \left( {t,x_1 ,x_2 ,x_3 } \right) = \left( {\sqrt {1 + q_2^2 } \sinh \left( {k\tau } \right),\cosh \left( {k\tau }
\right),q_2 \sinh \left( {k\tau } \right),0} \right)/k
\end{equation}
where $ q_2 $ is an ordinary signed scalar. Defining a new constant
\begin{equation} \label{13}
v_b  \equiv q_2 /\sqrt {1 + q_2^2 }
\end{equation}
(where evidently $ \left| {v_b } \right| < 1 $) one obtains
\begin{equation} \label{14}
x   = \frac{1} {k}\left( {\frac{{\sinh \left( {k\tau } \right)}} {{\sqrt {1 - v_b^2 } }},\cosh \left( {k\tau } \right),\frac{{v_b
\sinh \left( {k\tau } \right)}} {{\sqrt {1 - v_b^2 } }},0} \right)\,.
\end{equation}
Eliminating the proper time one has that the reaction-free trajectory is either of the branches of
\begin{equation} \label{15}
x_1    = \sqrt {\left( {1 - v_b^2 } \right)t^2  + 1/k^2 } ,\quad x_2   = v_b t\,.
\end{equation}
\section{Space-time geometry of the trajectory} \label{sec:SpaceTimeGeometry}
\subsection{General case} \label{sec:GeneralCase}
\begin{figure}
\begin{center}
\includegraphics*[scale = 1]{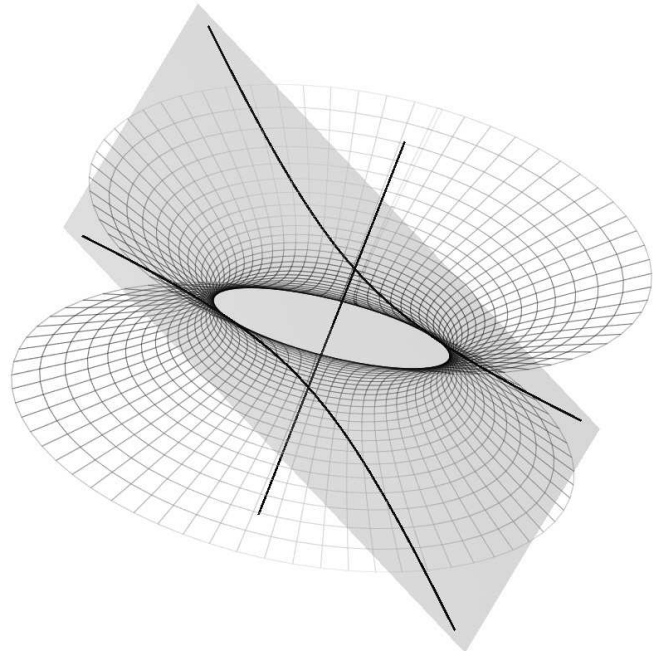}
\caption{\label{f1}
A reaction-free space-time trajectory in 1+2 D, depicted as the intersection of a plane with a
hyperboloid. The single heavy straight line is the time axis. The two heavy curved lines are the two branches of the
reaction-free hyperbolic space-time path given by Eq. \ref{15}. They can be regarded as the special case reaction-free space-time trajectories of Fig. \ref{f2} viewed from a frame moving at speed $v_b = 0.75$ in a direction normal to the shaded plane of that figure.}
\end{center}
\end{figure}

Eq. (15) describes the curve that is the intersection of the space-time plane
\begin{equation} \label{16}
x_2    = v_b t,\quad \left| {v_b } \right| < 1
\end{equation}
with the hyperboloid
\begin{equation} \label{17}
x_1^2  + x_2^2  = t^2  + 1/k^2\,.
\end{equation}
That is, equation (\ref{15}), for various $ v_b $ and $ k $, are sections of a space-time hyperboloid. Consequently the trajectories are hyperbolas in the sectioning plane and therefore are plane curves in two space dimensions and one time dimension. The space-time hyperboloid (\ref{17}) is always oriented along the time axis, asymptotic to a $45^o$ cone. It can be arbitrarily located in space and time, and arbitrarily oriented in 3D space. It can have any waist size (given by $1/k$). The sectioning space-time plane is arbitrary except for the constraint that the angle between its normal and the time axis, modulo $180^0$, must be greater than $45^o$. This guarantees that it cuts the hyperbola along a hyperbolic space-time path - it cannot cut the hyperboloid along an elliptical path. A particular case is depicted in Fig. \ref{f1}: the shaded plane is (\ref{16}) with $v_b = 0.75$, and the wire-frame surface is the hyperboloid (\ref{17}) (here with $k=1$).

From the considerations above it follows that the space projection of the hyperbolic path is entirely arbitrary. That is, a charge following \emph{any} hyperbolic path in 2 space dimensions can be rendered reaction-free - provided the speed on the path is chosen in accordance with (\ref{8}). For example, the spatial projection of (\ref{15}), i.e. the path through space without regard to the time, is the hyperbola
\begin{equation} \label{18}
x_1^2  - \left( {1/v_b^2  - 1} \right)x_2^2    = 1/k^2\,.
\end{equation}
This hyperbola (\ref{18}) is oriented along the $ x_1 $ axis and has major axis of length $ 1/k $ and asymptotes to the lines $ x_2 = \pm v_b x_1 / \sqrt {1 - v_b^2 } $,.

\begin{figure}
\begin{center}
\includegraphics*[scale = 1.4]{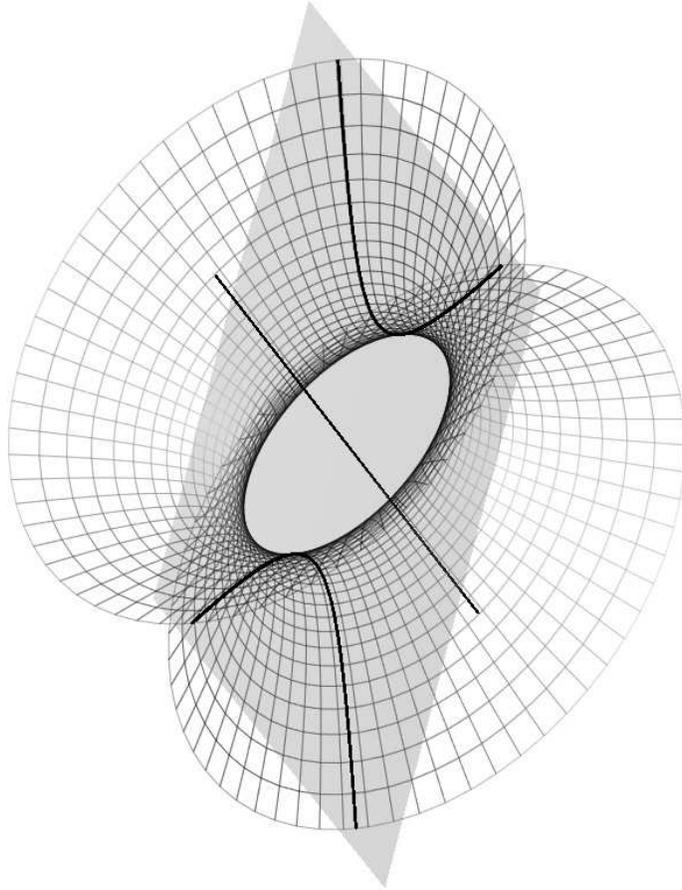}
\caption{\label{f2}
The traditional - special case - reaction-free space-time trajectory in 1+1 D, depicted as the
intersection of a plane containing the $t$-axis with a hyperboloid.}
\end{center}
\end{figure}

In the particular case that the sectional plane contains the time-axis, $ x_2  = 0 \Leftrightarrow v_b  = 0 $, one has from (\ref{16}) and (\ref{17}) that
\begin{equation} \label{19} x_2  = 0,\quad x_1  = \sqrt {t^2  + 1/k^2 }
\end{equation}
for the projection of the trajectory onto the $ t,x_1 $ axes. The corresponding space-only projection is then either of the semi-infinite straight lines $ x_1 > 1/k,\quad x_2  = 0 $, or $ x_1  <  - 1/k$, $x_2  = 0 $. This is the traditional case (or `special case') of a 1+1 D space-time hyperbola considered in the literature and is shown in Fig. \ref{f2}.

Viewing the special case trajectory from a moving frame traveling at speed $ v_b $ in the negative $ x_2 $ direction and referenced by primed coordinates, one has
\begin{equation} \label{20}
t = \frac{{t' - v_b x'_2 }} {{\sqrt {1 - v_b^2 } }},\quad x_1  = x'_1 ,\quad x_2  = \frac{{x'_2  - v_b t'}} {{\sqrt {1 - v_b^2 } }}\,.
\end{equation}
In the new coordinates (\ref{19}) becomes
\begin{equation} \label{21}
x'_1  = \sqrt {\frac{{\left( {t' - v_b x'} \right)^2 }} {{1 - v_b^2 }} + \frac{1} {{k^2 }}}  = \sqrt {\left( {1 - v_b^2 } \right)t'^2
+ 1/k^2 } ,\quad x'_2  = v_b t'
\end{equation}
which agrees with (\ref{15}). Hence it is clear that the novel degrees of freedom in the `non-special' space-time
trajectory (i.e., apart from the obvious freedoms of spatial orientation and space-time location) can be obtained from boosts of the special case hyperbolic space-time curve. This correspondence justifies the choice of the symbol $ v_b $, introduced in (\ref{13}); initially regarded as one of the arbitrary constants in the solution of (\ref{5}), it turns out to be the velocity of the boosted viewpoint of the special case. Additionally, one sees (retrospectively) that $q_2 $ in (\ref{13}) is the proper speed of the boosted viewpoint.

Rather than boosts of the special space-time curve, an alternative way to generate the family of reaction-free curves is to boost the special case surfaces - the hyperboloid and the sectioning plane - themselves. The hyperboloid (\ref{17}) is a special case of an invariant space-time surface under Lorentz boosts:
\begin{equation} \label{22}
x_1^2  + x_2^2  = t^2  + 1/k^2  \to x_1^{'2}  + x_2^{'2}  = t^{'2}  + 1/k^2\,.
\end{equation}
The plane $ x_2  = 0 $ is not an invariant surface, but transforms under boosts like
\begin{equation} \label{23}
x_2  = 0 \to x'_2  = v_b t'\,.
\end{equation}
With reference to for example Fig. \ref{f1}, the result now follows immediately that the family of reaction-free curves is generated by rotating the sectioning plane whilst leaving the hyperboloid unchanged.
\subsection{Uniform motion} \label{sec:UniformMotion}
\begin{figure}
\begin{center}
\includegraphics*[scale = 1]{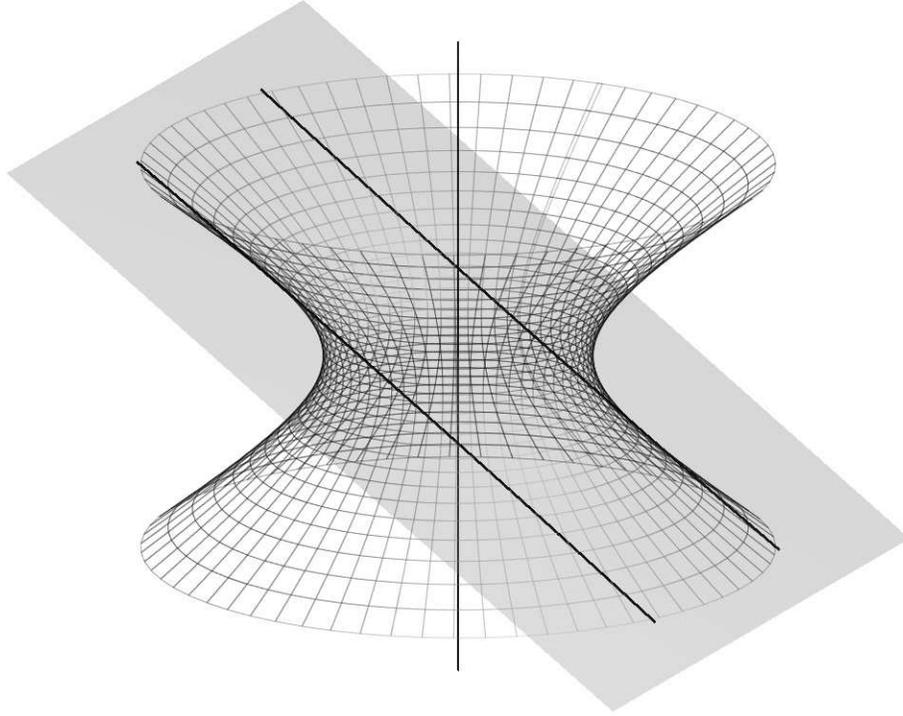}
\caption{\label{f3}
Parallel null rays resulting from a sectioning plane inclined at $45^o$ to the time axis. They
can be regarded as the special case trajectories of Fig. \ref{f1} viewed from a frame moving at light speed in a
direction normal to the shaded plane of that figure.}
\end{center}
\end{figure}

In the limit that the speed of the boosted viewpoint is $v_b = 1$, (\ref{16}) gives that the plane is inclined at
$45^o$ with respect to the time axis and Eq. (\ref{17}) then gives that $x_1 = \pm 1/k$. These reaction-free
trajectories are the two parallel straight line null rays in $x_2, t$ located at $x_1 = 1/k$ and $x_1 = -1/k$. Through reorientation and relocation of the axes and variation of the arbitrary value of $k$, the geometry can generate every possible pair of parallel null rays, a particular example of which is given in Fig. \ref{f3}.

The particular case that $k=0$ requires special treatment: From either (\ref{6}) or (\ref{7}) one has that there is no acceleration. Eq. (\ref{7}) then generates single trajectories with arbitrary velocity; rectilinear motion is
reaction-free.
\section{External force causing no reaction} \label{sec:ExtFNoReaction}
In the case that there is no reaction, all that remains of the Lorentz-Abraham-Dirac equation (\ref{1}) is that $
f_{ext}  = ma $ where $ f_{ext}  = \gamma \left( {{\mathbf{F}}{\mathbf{.v}},{\mathbf{F}}} \right) $ is the proper
4-force, and $\mathbf{F}$ is the ordinary, e.g. Lorentz, force. In the coordinate system located so that $ r = 0 $ one
has $ {\mathbf{a}} = k^2 {\mathbf{x}} $, and therefore from (\ref{14})
\begin{equation} \label{24}
{\mathbf{F}} = m{\mathbf{a}}/\gamma  = mk^2 {\mathbf{x}}/\gamma\,,
\end{equation}
where
\begin{equation} \label{25}
\gamma  = \sqrt {1 + {\mathbf{u}}^2 }    = \sqrt {1 + \sinh ^2 \left( {k\tau } \right) + \frac{{v_b^2 \cosh ^2 \left( {k\tau }
\right)}} {{1 - v_b^2 }}}  = \gamma _b \cosh \left( {k\tau } \right)
\end{equation}
where $ \gamma _b  = 1/\sqrt {1 - v_b^2 } $ . Note that the proper acceleration is not constant. Using this and
(\ref{14}) the components of force required to produce reaction-free motion are found to be
\begin{equation} \label{26}
F_1  = mk/\gamma _b ,\quad F_2  = v_b mk\tanh \left( {k\tau } \right) = \frac{{v_b mk^2 t}} {{\sqrt {\gamma _b^2  + k^2 t^2 } }}\,,
\end{equation}
where the ordinary time form of the last expression may be obtained from the 0th component of $x$ as given in (\ref{14}).

Clearly the special case $v_b  = 0$ requires only a constant ordinary force, for example a uniform electric field. If $
v_b  \ne 0 $ - corresponding to reaction-free trajectory that is hyperbolic in space - the force is still constant
along the major axis ($ x_1 $ -axis), whereas an additional transverse force is required that is odd in time (and
therefore in the direction of the minor axis). This component of force tends to the constant value $ F_2  \to v_b mk $
as $ \left| {x_2 } \right|,\left| t \right| \to \infty $. It may at first seem surprising that a transverse component of force is necessary, since the component of velocity of the charge in that direction is just the constant $ v_b $,
and therefore the ordinary acceleration in the direction of $ x_2 $ is zero. However the proper acceleration in that
direction is not zero; one has
\begin{equation} \label{27}
\frac{{d^2 {\mathbf{x}}}} {{d\tau ^2 }} = \gamma \frac{d} {{dt}}\left( {\gamma \frac{{d{\mathbf{x}}}} {{dt}}} \right) =
\gamma ^2 \frac{{d^2 {\mathbf{x}}}} {{dt^2 }} + \frac{1} {2}\frac{{d\gamma ^2 }} {{dt}}\frac{{d{\mathbf{x}}}} {{dt}}
\end{equation}
from which one observes that the proper acceleration in any fixed direction can be driven, via the term $ {{d\gamma ^2 }
\mathord{\left/ {\vphantom {{d\gamma ^2 } {dt}}} \right. \kern-\nulldelimiterspace} {dt}} $, by speed changes exclusively in other, orthogonal, directions.
\section{Geometric continuation} \label{sec:GeometricContinuation}
\subsection{Motivation} \label{subsec:Motivation}
Figures \ref{f1}, \ref{f2}, and \ref{f3} suggest an extrapolation, motivated entirely by the geometry, wherein the boost plane cuts the 2+1 D hyperboloid at an angle greater than 45 degrees to the vertical, thereby giving rise to a closed planar path. The strong geometric motivation for this extension is a by-product of the particular approach to the derivation above wherein all possible reaction-free space-time trajectories are derived from rotations of a sectioning plane. A purely algebraic analysis for example would very likely not generate the same conceptual momentum in favour of extrapolation. An immediate outcome of that extrapolation is that the motion is superluminal, though here this is regarded as a secondary consequence of the geometrically-motivated extrapolation.

With reference to Eqs. (\ref{16}) and (\ref{17}) the intersection of the plane
\begin{equation} \label{28}
x_2    = v_b t,\quad \left| {v_b } \right| > 1
\end{equation}
with the hyperboloid (\ref{17}) gives rise to the projection in the $x_1,x_2$ plane at
\begin{equation} \label{29}
x_1^2  + \left( {1 - 1/v_b^2} \right)x_2^2    = 1/k^2
\end{equation}
which in general is an ellipse - and a circle of radius $1/k$ in the limit that $v_b \rightarrow \infty$. These paths correspond to a superluminal boost $|v_b| > 1$ of the original purely 1+1 dimensional traditional hyperbolic path in Fig. \ref{f1}. Alternatively, they can be taken to represent the trajectories of superluminal charges viewed from the frame of Fig. \ref{f1}. Interpreted physically, a charge-point moving superluminally on a closed curved space-time path according to (\ref{28}) and (\ref{29}) is a pair creation and destruction event without radiation. In the limit that the trajectory is a circle, it represents a spatially extended event (or object) of zero (temporal) duration. In that limit there is no motion, just the appearance of a circular object of finite spatial extent.

A full investigation of possible \textit{physical} justifications for this extension of the mathematics and geometry to the superluminal domain is beyond the scope of this short article; in the following we will be content to make a few observations.

\subsection{Generalized proper time \\and Abraham - von Laue vector}  \label{sec:GeneralizedAVLvector}
The aim of this section is to find a generalized Abraham - von Laue vector compliant with the geometric intuition of Section \ref{subsec:Motivation}. It will be seen that it is sufficient just to extend the definition of proper time. From this a generalized Abraham - von Laue vector follows automatically, the vanishing of which generates the extended set of reaction-free trajectories.

The traditional definition of the proper time
\begin{equation} \label{30}
d\tau = \sqrt{dx^2}
\end{equation}
cannot parameterize a superluminal trajectory because the argument of the square root is then negative. The definition can be replaced with
\begin{equation} \label{30.5}
d\tau = \sqrt{|dx^2|}\
\end{equation}
to cover the explicit possibility of superluminal motion, so that $\tau$ increases monotonically along the trajectory regardless of the speed and the direction in $t$ time. The modulus operation has been found elsewhere to arise naturally in a formal extension of classical EM extended to accommodate superluminal motion, \cite{Ibison2,Ibison3}. It should be pointed out that the modulus arises in those analyses automatically - as if it were present all along, though is ignored in the sub-luminal domain because there it is redundant. In any case $d\tau = \sqrt{|dx^2|}$ satisfies the traditional criteria for a Lorentz scalar. That is, $\tau$ is unchanged by ordinary Lorentz transformations, including (sub-luminal) boosts. It does not matter that when the trajectory is superluminal the increments (\ref{30.5}) are \emph{space}-like rather than \emph{time}-like. It is important only that as a Lorentz scalar (\ref{30.5}) provides an invariant parameterisation of the trajectory.\footnote{Time-reversals, which are absent from the sub-luminal domain, are an essential feature of the superluminal reaction-free trajectories and must therefore be properly accommodated by the analysis. Though they entail no problems for the definition (\ref{30.5}) time reversals break the traditional correspondence between laboratory time and proper time. That is, the proper time remains a monotonically increasing parameter of the trajectory even when the laboratory time does not. For this reason the reader may object to the continued use of the term `proper time' to describe these generalized intervals and may have preferred the alternative `proper interval' because it does not suffer from this shortcoming. But then in order that it remain apparent that the definition (\ref{30.5}) applies to both domains it would be necessary to replace the more usual term `proper time' with `proper interval' in the sub-luminal domain also.} In fact, it is readily apparent that the time-like or space-like quality of the 4-interval $dx$ is not changed by a sub-luminal boost.

Henceforth the definition (\ref{30.5}) will now be understood to apply to a (now generalized) Lorentz velocity vector $u = dx/d\tau$. The parameterisation fails at light speed, but otherwise one then has
\begin{align*} \label{31.5}
u^2 =   & +1 \quad\text{if}\quad v<1 \\
        &-1 \quad\text{if}\quad v>1\,.
\end{align*}
With this and recalling the requirement that $u \circ \Gamma =0$ it follows that the generalized Abraham - von Laue vector must be
\begin{equation} \label{32}
\Gamma  =   \sigma\frac{{2e^2 }} {3} \left(\frac{da}{d\tau} +\text{sgn}\left(u^2\right) a^2 u \right)
\end{equation}
where $\sigma$ must be either $+1$ or $\text{sgn}\left(u^2\right)$. The new (internal) factor of $\text{sgn}\left(u^2\right)$ is necessary to guarantee that the 4-vector remains Lorentz-orthogonal to the proper velocity in both sub-luminal, and superluminal domains. The \emph{overall} sign ambiguity associated with $\sigma$ will have important consequences for the motion of the superluminal charge satisfying the full Lorentz Dirac equation, though that is not relevant to the focus of this document which is exclusively on reaction-free motion wherein the total Abraham - von Laue vector vanishes.
\section{Superluminal reaction-free motion I}  \label{sec:SuperI}
\begin{figure}
\begin{center}
\includegraphics*[scale = 0.85]{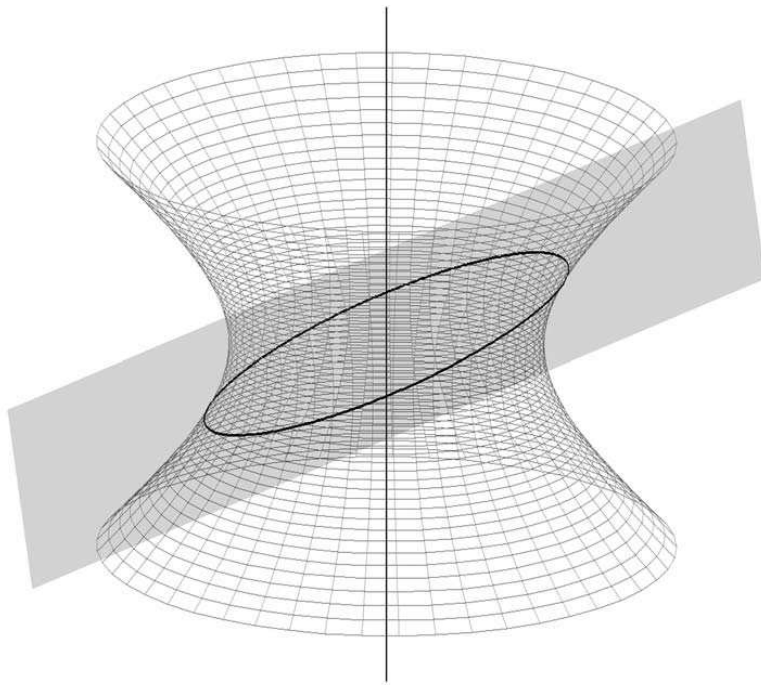}
\caption{\label{f4}
A closed path reaction-free trajectory. The ellipse may be considered as the result of a superluminal boost of a sub-luminal hyperbolic reaction-free trajectory.}
\end{center}
\end{figure}

\begin{figure}
\begin{center}
\includegraphics*[scale = 0.85]{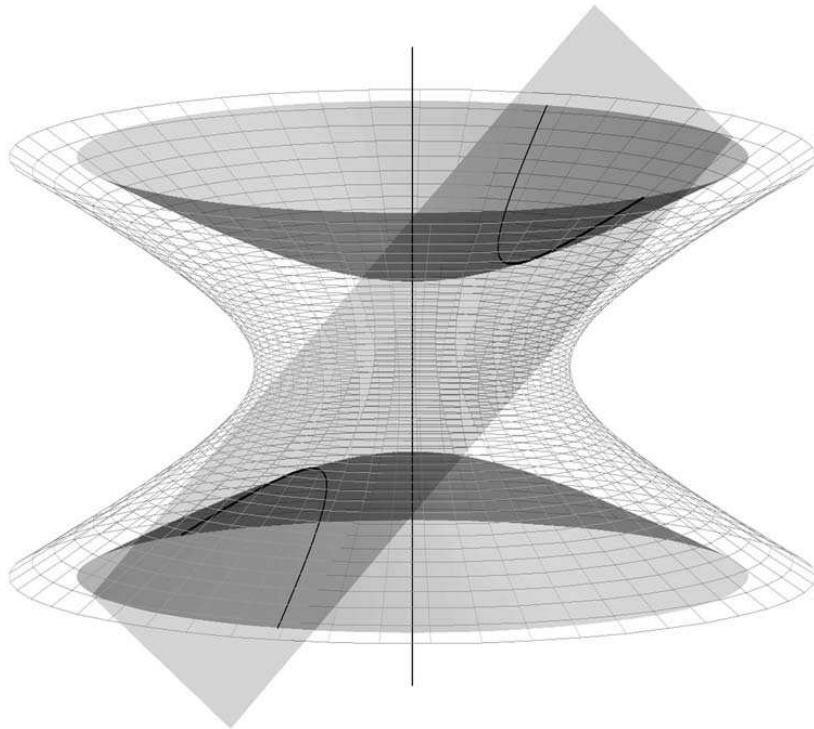}
\caption{\label{f5}
A pair of parabolic reaction-free trajectories. The lower trajectory is a pair destruction event in forward (laboratory) time, the upper trajectory is a pair creation event. The complete family of such trajectories is given by all possible orientations of the sectioning plane. All members are superluminal. (Note: at each orientation, the same sectioning plane gives rise to the sub-luminal hyperbolic reaction-free trajectories illustrated in Fig. \ref{f1}.)}
\end{center}
\end{figure}

In the superluminal domain there are two solutions to (\ref{32}) with $\Gamma=0$. One is
\begin{equation} \label{33}
x = r + \left(p\cos \left( {k\tau } \right) + q\sin \left( {k\tau } \right)\right)/k \,
\end{equation}
which requires
\begin{equation} \label{34}
p^2  =  q^2  = -1,\quad p \circ q = 0\,.
\end{equation}
It immediately follows that $a^2 = -k^2$, which ensures that (\ref{33}) is consistent with (\ref{32}). We now proceed much as in Section \ref{sec:ReactionFreeTrajectories}. Through choice of time origin one has
\begin{equation} \label{35}
p = \left( {0,{\mathbf{\hat p}}} \right),\quad q = \left( {\sqrt {{\mathbf{q}}^2 - 1} ,{\mathbf{q}}} \right),\quad {\mathbf{p}}{\mathbf{.q}} = 0\,.
\end{equation}
Again, since $\bf{p}$ and $\bf{q}$ are orthogonal, it is convenient to align $\bf{p}$ and $\bf{q}$ with the $ x_1 $ and $ x_2 $ axes respectively and locate the space and time axes so that $r = 0$. Then (\ref{33}) becomes
\begin{equation} \label{36}
x   = \left( {\sqrt {q_2^2 - 1} \sin \left( {k\tau } \right),\cos \left( {k\tau }\right),q_2 \sin \left( {k\tau } \right),0} \right)/k
\end{equation}
where $ q_2 $ is an ordinary signed scalar. Note that here the proper time is not monotonically related to the ordinary time. Defining a new constant
\begin{equation} \label{37}
v_b  \equiv q_2 /\sqrt {q_2^2 - 1 }
\end{equation}
(where evidently $ \left| {v_b } \right| > 1 $ ) one obtains
\begin{equation} \label{38}
x   = \frac{1} {k}\left( {\frac{{\sin \left( {k\tau } \right)}} {{\sqrt { v_b^2 - 1} }},\cos \left( {k\tau } \right),\frac{{v_b
\sin \left( {k\tau } \right)}} {{\sqrt {v_b^2 -1 } }},0} \right)\,.
\end{equation}
Eliminating the proper time, one has that the reaction-free trajectory is either of the branches of
\begin{equation} \label{39}
x_1    = \sqrt { 1/k^2 - \left( {v_b^2 -1} \right)t^2  } ,\quad x_2   = v_b t\,.
\end{equation}
Eq. (39) describes the curve that is the intersection of the space-time plane
\begin{equation} \label{40}
x_2    = v_b t,\quad \left| {v_b } \right| > 1
\end{equation}
with the hyperboloid
\begin{equation} \label{41}
x_1^2  + x_2^2  = t^2  + 1/k^2\,.
\end{equation}
Eliminating the time in favor of $x_1$ and $x_2$, one recovers the expected (\ref{29}), so that the trajectories are ellipses in the sectioning plane, becoming a circle in the limit of infinite boost $v_b$, depicted by Figs. \ref{f4} and \ref{f5} respectively. This confirms that the reaction free solutions of Eq. (\ref{32}) correspond to the geometric continuation from the sub-luminal domain of reaction-free trajectories discussed above. Accordingly, Eq. (\ref{32}) is a good candidate for a generalized Abraham - von Laue vector. Given this generalization, the sectioning plane can be rotated without restriction through $[0,\pi]$, each intersection with the hyperboloid (\ref{41}) being a space-time planar curve that is reaction-free.

\section{External force causing no reaction}  \label{sec:SuperExtFNoReaction}
As for the sub-luminal case it is interesting to determine the external force required to cause no reaction. Adopting the parameterisation (\ref{30.5}), the generalization of the reaction-free version of the sub-luminal Lorentz-Dirac equation can differ at most by a sign from its subluminal form. In particular, in the case of an external EM force one must have
\begin{equation} \label{41.2}
m\frac{d^2 x^a} {d\tau ^2 } = \pm F^{ab}\frac{d x_b} {d\tau }\,.
\end{equation}
Putting in from (\ref{38}), it can be shown that, up to the unknown sign, a sufficient - but not most general - set of EM fields is
\begin{equation} \label{41.4}
\mathbf{E} = \left( 0 ,\frac{ m k}{e \sqrt{\left(v_b^2-1\right)}},0\right)\,,\quad \mathbf{B} = \left( -\frac{v_b m k}{e \sqrt{\left(v_b^2-1\right)}},0,0\right)\,.
\end{equation}
When the sectioning plane is at constant $t$, $v_b$ is infinite, the reaction-free trajectory is a circle at constant time, and
\begin{equation} \label{41.6}
\mathbf{E} = \mathbf{0},\quad \mathbf{B} = -\left( m k/e ,0,0\right)\,,
\end{equation}
a result that establishes a duality correspondence between the sub-and super-luminal cases; in the former case a reaction-free trajectory in 1+1D can be sustained with a constant electric field, in the latter case a constant magnetic field will sustain a reaction-free trajectory in 2 space dimensions (with zero temporal extent). In both domains the more general cases can be obtained from (sub-luminal) Lorentz transformation of these special cases whereupon the transformed trajectories occupy one time and two space dimensions, and the transformed Faraday tensor contains non-zero contributions from both electric and magnetic fields.
\section{Superluminal reaction-free motion II}  \label{sec:SuperII}
Eq. (\ref{32}) admits a second solution
\begin{equation} \label{42}
x = r + \left(q\cosh \left( {k\tau } \right) + p\sinh \left( {k\tau } \right)\right)/k \,
\end{equation}
and therefore
\begin{equation} \label{43}
u = q\sinh \left( {k\tau } \right) + p\cosh \left( {k\tau } \right) \,
\end{equation}
which requires
\begin{equation} \label{44}
q^2  =  -p^2  = 1,\quad p \circ q = 0\,.
\end{equation}
The acceleration is
\begin{equation} \label{45}
a = k\left(q\cosh \left( {k\tau } \right) + p\sinh \left( {k\tau } \right)\right) \,
\end{equation}
and therefore $a^2 = k^2$, which ensures that (\ref{42}) is consistent with (\ref{32}).
\begin{equation} \label{46}
 p= \left( {0,{\mathbf{\hat p}}} \right),\quad q = \left( {\sqrt {{\mathbf{q}}^2 + 1} ,{\mathbf{q}}} \right),\quad {\mathbf{p}}{\mathbf{.q}} = 0\,.
\end{equation}
Again, since $\bf{p}$ and $\bf{q}$ are orthogonal, it is convenient to align $\bf{p}$ and $\bf{q}$ with the $ x_1 $ and $ x_2 $ axes respectively, and locate the space and time axes so that $r = 0$. Then (\ref{42}) becomes
\begin{equation} \label{47}
x   = \left( {\sqrt {q_2^2 + 1} \cosh \left( {k\tau } \right),\sinh \left( {k\tau }\right),q_2 \cosh \left( {k\tau } \right),0} \right)/k
\end{equation}
where $ q_2 $ is an ordinary signed scalar. Defining a new constant
\begin{equation} \label{48}
v_b  \equiv q_2 /\sqrt {q_2^2 + 1 }
\end{equation}
(where evidently $ \left| {v_b } \right| < 1 $ ) one obtains
\begin{equation} \label{49}
x   = \left( \frac{1}{\sqrt{1-v_b^2}} \cosh \left( {k\tau } \right),\sinh \left( {k\tau }\right),\frac{v_b}{\sqrt{1-v_b^2}} \cosh \left( {k\tau } \right),0 \right)/k\,.
\end{equation}
Eq. (49) describes the curve that is the intersection of the space-time plane
\begin{equation} \label{50}
x_2    = v_b t,\quad \left| {v_b } \right| < 1
\end{equation}
with the paraboloid
\begin{equation} \label{51}
x_1^2  + x_2^2  = t^2  - 1/k^2\,.
\end{equation}
Eliminating the time in favor of $x_1$ and $x_2$
\begin{equation} \label{52}
x_2^2\left(1/v_b^2-1\right) = x_1^2 + 1/k^2
\end{equation}
the trajectory is seen to be a parabola (in $\left(x_1,x_2\right)$).

It can be shown that the force required to sustain this superluminal (II) reaction-free orbit is the same as for case I except for a change in sign.

\section{Discussion}  \label{sec:Interpretation}
\subsection{Further remarks on the geometry}  \label{sub:geometry}
It is clear from the geometry that all reaction-free trajectories are conic sections in space-time. It follows that these trajectories could be presented exactly that way, i.e. as the loci of intersection between a plane and a 2+1 D cone (suppressing 1 space-dimension). Generally however, the 2+1 D cone in question is not a light cone, and therefore does not transform into itself under boosts. Consequently such a representation does not permit a straightforward geometrical interpretation of the relationship between the members of the complete family of reaction-free trajectories. By contrast, with the reaction-free trajectories represented (here) as intersections of a plane with a 2+1 D hyperboloid, all family members are easily seen as related by Poincar\'{e} transformations of the frame of reference in which the trajectory is the `traditional' 1+1 D hyperbolic path, simply by Poincar\'{e} transformations of the sectioning plane alone.

\subsection{Radiation again}  \label{sub:RadiationAgain}
The debate on whether or not the reaction-free trajectories are also free of radiation turns on the alleged impossibility - or at least the impracticality - of perfect hyperbolic motion for all time. Those concerns do not seem relevant to the elliptical closed path superluminal trajectories discussed in Sections \ref{sec:SuperI} and \ref{sec:SuperII}. Assuming so, it is interesting that these trajectories satisfy the observational requirement of stability that classical theory was (historically) deemed unable to supply, and which shortcoming was responsible for its replacement by quantum theory. Since these orbits are net charge-neutral, it seems unlikely, however, that they can be relevant to that problem.

\subsection{Pair creation}  \label{sec:PairCreation}
As the plane of simultaneity for the forwards in time traveling observer meets a closed spacetime trajectory described in Section \ref{sec:SuperI}, that trajectory will appear as the collision between two particles, one traveling forwards and the other backwards in time. Granted CPT invariance, the particle traveling backwards in time can be re-interpreted as an oppositely-charged particle traveling forwards in time, with the directions of space inverted. Therefore, as the plane of simultaneity makes a tangent to the worldline of a time-reversing particle, i.e. at the point that it changes direction in time, that event can be re-interpreted as pair-creation or pair-destruction.

Classical CPT invariance follows from the definition of the classical 4-current:
\begin{equation} \label{eqn:4current}
j\left( {x;e} \right) = e\int {d\lambda \frac{{dz\left( \lambda  \right)}}
{{d\lambda }}} {\delta ^4}\left( {x - z\left( \lambda  \right)} \right) = e\int {dz\left( \lambda  \right)} {\delta ^4}\left( {x - z\left( \lambda  \right)} \right)\,.
\end{equation}
Then
\begin{equation} \label{eqn:TR4current}
\begin{aligned}
j\left( { - x;e} \right) &= e\int {dz\left( \lambda  \right)} {\delta ^4}\left( {x + z\left( \lambda  \right)} \right) \\&= - e\int {d\tilde z\left( \lambda  \right)} {\delta ^4}\left( {x - \tilde z\left( \lambda  \right)} \right) \\&= j\left( {x; - e} \right)
\end{aligned}
\end{equation}
where $\tilde z\left( \lambda  \right) =  - z\left( \lambda  \right)$. That is, the classical current is invariant under simultaneous charge, parity, and time reversals. Note that the form of (\ref{eqn:4current}) guarantees charge conservation at all laboratory times, independent of the direction in time of the segments of the trajectory.

\subsection{A classical matter vacuum}  \label{sec:VacuumDecay}
The extended classical theory described above permits spontaneous classical pair-creation in the presence of a static EM field - a static magnetic field in the case of perfectly circular motion. No radiation is associated with these closed trajectories. Further, it can be deduced from the geometry that no part of a trajectory is electromagnetically visible to any other. That is, these trajectories are free of self-interaction except locally at the instantaneous charge-point (see for example \cite{Ibison3}), despite the fact they are superluminal and therefore generally vulnerable to non-local self-action. Consequently each closed trajectory can be considered, to first order, as a spontaneous pair creation and destruction event in the presence of the static field.

Recalling (\ref{41.6}) and (\ref{51}), the hyperboloid for the pair creation and destruction process has an interior radius that goes to infinity as the magnitude of the external field goes to zero. In the limit of zero external field the world lines for the pairs are straight, parallel, and light-like (null). Since these are the only constraints it follows that any oppositely charged pair of parallel null currents passing through \emph{any} location at \emph{any} orientation is a `solution' of the `vacuum' state. `Vacuum' here refers to the absence of an external driving field. The energy associated with this pair would ordinarily be presumed to have a mechanical and electromagnetic part, though at light speed neither can be decided unambiguously by extrapolation from the traditional sub-luminal form
\begin{equation} \label{eqn:massAction}
I  =  - m_0 \int {dt} \sqrt {1 - {\mathbf{v}}^2 }\,.
\end{equation}
The reader is referred to \cite{Ibison3} for an analysis leading to an argument for a particular resolution of this ambiguity in which the total energy of a classical point charge is zero precisely at light speed, and finite with either sign for infinitesimal departures from light speed.\footnote{The deliberations in \cite{Ibison3} are concerned only with the correct form of a time-symmetric self-action for a charge at or near light speed, whereas the reaction force under discussion here can be attributed exclusively to retarded self-action. A proper treatment requires the two approaches be integrated.}

Borrowing from those results, it follows that a Lorentz-Dirac equation extended to the superluminal domain gives rise to a classical vacuum populated by any number of light-like charge-pairs in rectilinear motion, free of radiation and self reaction. (This vacuum is field-free only to first order, neglecting interaction within and between charge pairs.) Further, in \cite{Ibison3} it is shown that the energy in the superluminal domain is unbounded from below, so this approach predicts that an empty vacuum is unstable and will spontaneous decay via pair creation.

Clearly there are some pointers here suggestive of a correspondence with the Dirac sea of the first quantized theory, and with the vacuum state of the corresponding field theory. But there is no prescription in this approach to determine the phase-space distribution of vacuum states. A similar situation arises in the sub-field of classical EM known as Stochastic Electrodynamics (SED) - see for example \cite{Marshall,BoyerReview} and \cite{Milonni} for a review - in which the standard Maxwell theory is supplemented by a classical background vacuum field that solves the homogeneous Maxwell equations with the constraint that it mimic the electromagnetic ZPF of the second quantized theory. This `classical ZPF' is homogeneous, isotropic, and Lorentz Invariant \cite{IbisonHaisch, BoyerSciAm} and reproduces at least to second order the statistics of the vacuum state electric and magnetic fields of QED. Calculations performed within SED correctly predict the Casimir effect (the typical calculation in QED such as that by Itzykson and Zuber \cite{IZ} is essentially classical), the Davies-Unruh effect \cite{BoyerFoRT}, and the van der Waals force \cite{BoyerReview,BoyerThermo}.

Borrowing from the SED tradition, the phase-space distribution of classical vacuum currents could be chosen to mimic, as far as possible, the classical observables associated with the Fermionic QFT vacuum, consistent with the restrictions on the current that it be composed of oppositely charged parallel and light-like trajectories. For example, the resulting classical vacuum current might be written
\begin{equation} \label{eqn:vacuumCurrent}
\begin{aligned}
  j\left(x\right) \equiv  & \left( {\rho \left( {t,{\mathbf{x}}} \right),{\mathbf{j}}\left( {t,{\mathbf{x}}} \right)} \right) \\
   =  & e\mathop{{\int\!\!\!\!\!\int}\mkern-21mu \bigcirc}
 {d{\Omega _{\mathbf{v}}}} \iiint {{d^3}a}\left( {1,{\mathbf{\hat v}}} \right)\xi \left( {{\mathbf{a}},{\mathbf{\hat v}}} \right){\delta ^3}\left( {{\mathbf{x}} - {\mathbf{a}} - {\mathbf{\hat v}}t} \right) \\
   =  & e\mathop{{\int\!\!\!\!\!\int}\mkern-21mu \bigcirc}
 {d{\Omega _{\mathbf{v}}}} \left( {1,{\mathbf{\hat v}}} \right)\xi \left( {{\mathbf{x}} - {\mathbf{\hat v}}t,{\mathbf{\hat v}}} \right) \\
\end{aligned}
\end{equation}
where $\xi \left( {{\mathbf{a}},{\mathbf{\hat v}}} \right)$ is a 4-vector random variable representing the net (signed) current density at time $t = 0$ passing though point $\mathbf{x}=\mathbf{a}$ in direction ${\mathbf{\hat v}}$, per unit phase space (per unit volume of $\mathbf{a}$ space and per unit solid angle of ${\mathbf{\hat v}}$ space). With this, the vacuum total current $j\left( x \right)$ is now a classical random variable.

A full treatment along these lines would follow the steps in \cite{IbisonHaisch} computing, say, the distribution function for the 4-current in the vacuum state of the Dirac field. At the least, for homogeneous neutrality of the vacuum, the statistical properties of $\xi \left( {{\mathbf{a}},{\mathbf{\hat v}}} \right)$ should be such that $\left\langle {j\left(x \right)} \right\rangle  = 0$. And for invariance of the vacuum at the level of covariance of the components of the current could choose the correlation statistics of $\xi \left( {{\mathbf{a}},{\mathbf{\hat v}}} \right)$ so that
\begin{equation}
\left\langle {{j_a}\left( x \right){j_b}\left( {x'} \right)} \right\rangle  = {g_{ab}}f\left( {{{\left( {x - x'} \right)}^2}} \right)
\end{equation}
where $g$ is the Minkowski metric and $f$ is some scalar function. 

It should be kept in mind however that these are classical distributions of classical currents. Accordingly, it may be that they can play a role analogous to that of the classical EM ZPF in SED, though this remains to be demonstrated. If so, it would imply the possibility of classical explanations for, say, the Fermionic Casimir effect \cite{Milton} and / or vacuum polarization.

\section{Acknowledgements}
The author is grateful to Mario Rabinowitz for interesting enjoyable exchanges and some helpful comments.



\begin{thebibliography}{}
\bibitem{RohrlichAmJoP} Rohrlich, F.: The self-force and radiation-reaction. Am. J. Phys. 68, 1109-1112 (2000).

\bibitem{RohrlichSelfForce} Rohrlich, F.: Classical self-force. Phys. Rev. D 60, 084017-1 - 084017-5 (1999).

\bibitem{Erber} Erber, T.: The Classical theories of radiation reaction. Fortschritte der Physik 9, 343-342 (1961).

\bibitem{Born} Born, M.: Ann. Phys. Lpz. 30, 1 (1909).

\bibitem{Pauli} Pauli, W.: Theory of relativity. Dover, New York (1981).

\bibitem{Bondi} Bondi, H. \& Gold, T.: The field of a uniformly accelerated charge, with special reference to the problem of gravitational acceleration. Proc. Roy. Soc. Lon. A 229, 416-424 (1955).

\bibitem{Fulton} Fulton, T. \& Rohrlich F.: Classical radiation from a uniformly accelerated charge. Annals of Physics 9, 499-517 (1960).

\bibitem{Boulware} Boulware, D.G.: Radiation from a uniformly accelerated charge. Annals of Physics 124, 169-188 (1979).

\bibitem{Bradbury} Bradbury, T.C.: Radiation damping in classical electrodynamics. Annals of Physics 19, 323-347 (1962).

\bibitem{Drukey} Drukey, D.L.: Radiation from a uniformly accelerated charge. Phys. Rev. 76, 543-544 (1949).

\bibitem{Singal1} Singal, A.K.: The Equivalence principle and an electric charge in a gravitational field. Gen. Rel. \& Grav. 27, 953-967 (1995).

\bibitem{Singal2} Singal, A.K.: The Equivalence Principle and an electric charge in a gravitational field II. A uniformly accelerated charge does not radiate. Gen. Rel. \& Grav. 27, 1371-1390 (1997).

\bibitem{Parrott1} Parrott, S.: Radiation from a charge uniformly accelerated for all time. Gen. Rel. \& Grav. 29, 1463-1472 (1997).

\bibitem{Parrott2} Parrott, S.: Radiation from a uniformly accelerated charge and the equivalence principle. Found. Phys. 32, 407-440 (2002).

\bibitem{Peierls} Peierls, R.: Relativity. In: Peierls, R. Surprises in theoretical physics, pp. 160-166. Princeton University Press, Princeton, NJ (1979).

\bibitem{Parrottbook} Parrott, S.: Relativistic electrodynamics and Differential Geometry. Springer, New York (1987).

\bibitem{RohrlichBook} Rohrlich, F.: Classical charged particles : foundations of their theory. Addison-Wesley, Redwood City, CA, (1990).

\bibitem{Ibison1} Ibison, M. \& Puthoff H.E.: Relativistic integro-differential form of the Lorentz-Dirac equation in 3D without runaways. J. Phys. A. 34, 3421-3428 (2001).

\bibitem{FordEtAl} Ford, G. W. \& O'Connell, Phys. Lett. A 157, 217 (1991).

\bibitem{Jackson} Jackson, J.D.: Classical electrodynamics. Wiley, New York (1998).

\bibitem{Spohn} Spohn, H.: The critical manifold of the Lorentz-Dirac equation. Europhys. Lett. 50,  287-292 (2000).

\bibitem{RohrlichCorrectEquation} Rohrlich, F.: The correct equation of motion of a classical point charge. Phys. Lett. A 283, 276-278 (2001).

\bibitem{MonizSharp} Moniz, E. J. \& Sharp, D. H.: Radiation reaction in nonrelativistic quantum electrodynamics. Phys. Rev. D 15, 2850-2865 (1977).

\bibitem{BarutZanghi} Barut, A. O. \& Zanghi, N.: Classical model of the Dirac Electron. Phys. Rev. Lett. 52, 2009-2012 (1984).

\bibitem{Medina} Medina, R.: Radiation reaction of a classical quasi-rigid extended particle. J. Phys. A 39, 3801-3816 (2006).

\bibitem{Ibison2} Ibison, M.: Un-renormalized classical electromagnetism. Annals of Physics 321, 261-305 (2006).

\bibitem{Ibison3} Ibison, M.: A new case for direct action. To appear in: Proc. Physical Interpretations of Relativity Theory 2008 (PIRT XI). Arxiv: 0810.4618.

\bibitem{Marshall} Marshall, T. W.: Random Electrodynamics. Proc. R. Soc. Lond. A 276, 475 (1963).

\bibitem{BoyerReview} Boyer, T. H.: Random electrodynamics: The theory of classical electrodynamics with classical electromagnetic zero-point radiation. Phys. Rev. D 11, 790-808 (1975).

\bibitem{Milonni} Milonni, P. W.: The Quantum Vacuum: An Introduction to Quantum Electrodynamics. Academic Press (1993).

\bibitem{IbisonHaisch} Ibison, M. \& Hasich, B.: Quantum and classical statistics of the electromagnetic zero-point field. Phys. Rev. A 54, 2737-2744 (1996).

\bibitem{BoyerSciAm} Boyer, T. H.: The classical vacuum. Scientific American 253, 56 (1985).

\bibitem{IZ} Itzykson, C. \& Zuber, J.-B.: Quantum field theory, McGraw-Hill, New York (1985).

\bibitem{BoyerFoRT} Boyer, T. H.: Chap. 5 in Foundations of Radiation Theory and Quantum
Electrodynamics, 1st ed., Edited by A. O. Barut. Plenum Press, New York (1980).

\bibitem{BoyerThermo} Boyer, T. H.: Classical statistical thermodynamics and electromagnetic zero-point
radiation. Phys. Rev. 180, 19 (1969).

\bibitem{Milton} Milton, K. A.: The Casimir Effect: Physical manifestations of zero-point energy, World Scientific, Hackensack, NJ (2001).

\end{thebibliography}
\end{document}